
%
\input harvmac

\def\Titlehh#1#2{\nopagenumbers\abstractfont\hsize=\hstitle\rightline{#1}%
\vskip .2in\centerline{\titlefont #2}\abstractfont\vskip .2in\pageno=0}
\def\CTPa{\it Center for Theoretical Physics, Department of Physics,
      Texas A\&M University}
\def\CTPb{\it College Station, TX 77843-4242, USA}
\def\HARCa{\it Astroparticle Physics Group,
Houston Advanced Research Center (HARC)}
\def\HARCb{\it The Woodlands, TX 77381, USA}
\def\UAa{\it Department of Physics and Astronomy, University of Alabama}
\def\UAb{\it Box 870324, Tuscaloosa, AL 35487-0324, USA}
\def\CERN{\it CERN Theory Division, 1211 Geneva 23, Switzerland}
\def\ie{\hbox{\it i.e.}}     
\def\eg{\hbox{\it e.g.}}

\def\coeff#1#2{{\textstyle{#1\over #2}}}

\catcode`\@=11 

\def\lsim{\mathrel{\mathpalette\@versim<}}
\def\gsim{\mathrel{\mathpalette\@versim>}}
\def\@versim#1#2{\vcenter{\offinterlineskip
    \ialign{$\m@th#1\hfil##\hfil$\crcr#2\crcr\sim\crcr } }}
\def\boxit#1{\vbox{\hrule\hbox{\vrule\kern3pt
      \vbox{\kern3pt#1\kern3pt}\kern3pt\vrule}\hrule}}

\def\etal{{\it et. al.}}

\def\t1{{\tilde 1}}

\def\JL{J. L. Lopez}
\def\DVN{D. V. Nanopoulos}

\def\eV{\,{\rm eV}}

\def\GeV{\,{\rm GeV}}
\def\TeV{\,{\rm TeV}}

\def\NPB#1#2#3{Nucl. Phys. B {\bf#1} (19#2) #3}
\def\PLB#1#2#3{Phys. Lett. B {\bf#1} (19#2) #3}

\def\PRD#1#2#3{Phys. Rev. D {\bf#1} (19#2) #3}
\def\PRL#1#2#3{Phys. Rev. Lett. {\bf#1} (19#2) #3}
\def\PRT#1#2#3{Phys. Rep. {\bf#1} (19#2) #3}
\def\MODA#1#2#3{Mod. Phys. Lett. A {\bf#1} (19#2) #3}

\def\TAMU#1{Texas A \& M University preprint CTP-TAMU-#1}
\def\ARAA#1#2#3{Ann. Rev. Astron. Astrophys. {\bf#1} (19#2) #3}
\def\ARNP#1#2#3{Ann. Rev. Nucl. Part. Sci. {\bf#1} (19#2) #3}

\nref\Smoot{G. F. Smoot, \etal, COBE preprint (1992).}
\nref\dvn{For a review see, \DVN, in Proceedings of the International
School of Astroparticle Physics, HARC January 1991, ed. \DVN\ (World
Scientific 1991), p. 183.}
\nref\Wright{E. L. Wright, \etal, COBE preprint (1992); R. L. Davis,
H. M. Hodges, G. F. Smoot, P. J. Steinhardt, and M. S. Turner, Fermilab
preprint FERMILAB-PUB-92/168-A.}
\nref\inflation{For recent reviews see \eg, K. Olive, \PRT{190}{90}{307};
D. Goldwirth and T. Piran, \PRT{214}{92}{223}.}
\nref\CDM{For a recent review see M. Davis, G. Efstathiou, C. S. Frenk,
and S. D. M. White, {\it Nature} {\bf356} (1992) 489.}
\nref\NO{See \eg, \DVN\ and K. Olive, Nature {\bf327} (1987) 487.}
\nref\many{R. Schaefer and Q. Shafi, Bartol preprint BA-92-28 (1992);
G. Efstathiou, J. R. Bond, and S. D. M. White, Oxford University preprint
OUAST/92/11 (1992); A. N. Taylor and M. Rowan-Robinson, Queen Mary College
preprint, (1992); M. Davis, F.J. Summers and D. Schlegel, Berkeley preprint
CfPA-TH-92-016 (1992).}
\nref\DM{For reviews on dark matter and its detection, see \eg, V. Trimble,
\ARAA{25}{87}{425}; J. R. Primack, B. Sadoulet, and D. Seckel,
\ARNP{38}{88}{751}; D. O. Caldwell, \MODA{5}{90}{1543}; P. F. Smith and J. D.
Lewin, \PRT{187}{90}{203}.}
\nref\EHNOS{H. Goldberg, \PRL{50}{83}{1419}; J. Ellis, J. S. Hagelin, \DVN,
K. A. Olive, and M. Srednicki, \NPB{238}{84}{453}.}
\nref\usualdm{J. Ellis, J. Hagelin, and D. V. Nanopoulos, \PLB{159}{85}{26};
K. Griest, \PRD{38}{88}{2357}; \PRD{39}{89}{3802}(E);
K. A. Olive and M. Srednicki, \PLB{230}{89}{78} and \NPB{355}{91}{208};
K. Griest, M. Kamionkowski, and M. S. Turner, \PRD{41}{90}{3565};
J. Ellis, L. Roszkowski, and Z. Lalak, \PLB{245}{90}{545};
L. Roszkowski, \PLB{262}{91}{59}; K. Griest and L. Roszkowski, Berkeley
preprint CfPA-Th-91-013; L. Roszkowski, CERN preprint CERN-TH.6264/91;
J. McDonald, K. A. Olive, and M. Srednicki, \PLB{283}{92}{80}.}
\nref\EZ{J. Ellis and F. Zwirner, \NPB{338}{90}{317}.}
\nref\Nojiri{M. M. Nojiri, \PLB{261}{91}{76}.}
\nref\LNYa{\JL, \DVN, and K. Yuan, \NPB{370}{92}{445}.}
\nref\LNYb{\JL, \DVN, and K. Yuan, \PLB{267}{91}{219}.}
\nref\ER{J. Ellis and L. Roszkowski, \PLB{283}{92}{252}.}
\nref\noscale{S. Kelley, \JL, \DVN, H. Pois, and K. Yuan, \PLB{273}{91}{423}.}
\nref\DNdm{M. Dress and M. M. Nojiri, SLAC preprint SLAC-PUB-5860 (July 1992).}
\nref\aspects{S. Kelley, \JL, \DVN, H. Pois, and K. Yuan, \TAMU{16/92} and
CERN-TH.6498/92.}
\nref\ERZ{J. Ellis, G. Ridolfi, and F. Zwirner, \PLB{257}{91}{83};
M. Drees and M. M. Nojiri, \PRD{45}{92}{2482}.}
\nref\higgs{S. Kelley, \JL, \DVN, H. Pois, and K. Yuan, \PLB{285}{92}{61}.}
\nref\Griest{K. Griest and D. Seckel, \PRD{43}{91}{3191};
P. Gondolo and G. Gelmini, \NPB{360}{91}{145}.}
\nref\troubles{\JL, \DVN, and A. Zichichi, \TAMU{49/92} and CERN-TH.6554/92;
\JL, \DVN, and H. Pois, in preparation.}

\nfig\I{The neutralino relic density $\Omega_\chi h^2_0$ in the $(M_2,\mu)$
plane for $\tan\beta=2,8$, $\xi_A=0$, both signs of $\mu$, and
(a) $\xi_0=0$, (b) $\xi_0=0.4$, (c) $\xi_0=1.0$. The various symbols denote
the following ranges of $\Omega_\chi h^2_0$: $<0.025$ (dots), $0.025-0.10$
(crosses), $0.10-0.25$ (diamonds), $0.25-0.50$ (squares), $0.50-1.0$ (pluses),
$>1.0$ (stars). For $h=1,1/\sqrt{2},0.5$
stars,stars+pluses,stars+pluses+squares are excluded on cosmological grounds.}
\nfig\II{The neutralino relic density $\Omega_\chi h^2_0$ as a function of the
gluino mass $m_{\tilde g}$ along rays with $m_t=150\GeV$ in Fig. 1c
($\xi_0=1,\xi_A=0$) and Fig. 3 ($\xi_0=2,\xi_A=0$). (In Fig. 1c these rays
correspond to $M_2/|\mu|\approx0.5\,(0.7)$ for $\tan\beta=2\,(8)$.) Solid
(dashed) lines correspond to $\mu>0$ ($\mu<0$). Note the various poles and
thresholds of the neutralino annihilation cross section.}
\nfig\III{The neutralino relic density $\Omega_\chi h^2_0$ in the $(M_2,m_t)$
plane for $\xi_0=2,\xi_A=0$, for $\tan\beta=2,8$ and both signs of $\mu$.
The symbols are as in Fig. 1. For $h=1,1/\sqrt{2},0.5$
stars,stars+pluses,stars+pluses+squares are excluded on cosmological grounds.}
\nfig\IV{The fraction of the allowed parameter space in the $(M_2,m_t)$ plane
which is cosmologically excluded for $h=1/2$ ($f_{h=1/2}$) versus $\xi_0$
for $\xi_A=-\xi_0,0,+\xi_0$, both signs of $\mu$ and $\tan\beta=$ 2 (solid),
8 (dashed), 15 (dotdashed), 30 (dotted).}

\Titlehh{\vbox{\baselineskip12pt\hbox{CERN-TH.6584/92}\hbox{CTP--TAMU--56/92}
\hbox{ACT--16/92}\hbox{UAHEP9212}}}
{\vbox{\centerline{New Constraints on Neutralino Dark Matter}
        \vskip2pt\centerline{in the Supersymmetric Standard Model}}}
\centerline{S. KELLEY$^{(a)(b)}$,
JORGE~L.~LOPEZ$^{(a)(b)}$, D.~V.~NANOPOULOS$^{(a)(b)(c)}$,}
\centerline{H. POIS$^{(a)(b)}$, and KAJIA YUAN$^{(d)}$}
\smallskip
\centerline{$^{(a)}$\CTPa}
\centerline{\CTPb}
\centerline{$^{(b)}$\HARCa}
\centerline{\HARCb}
\centerline{$^{(c)}$\CERN}
\centerline{$^{(d)}$\UAa}
\centerline{\UAb}
\vskip .1in
\centerline{ABSTRACT}
We investigate the prospects for neutralino dark matter within the
Supersymmetric Standard Model (SSM) including the constraints from
universal soft supersymmetry breaking and radiative breaking of the electroweak
symmetry. The latter is enforced by using the one-loop Higgs effective
potential which automatically gives the one-loop corrected Higgs boson masses.
We perform an exhaustive search of the allowed five-dimensional parameter space
and find that the neutralino relic abundance $\Omega_\chi h^2_0$ depends most
strongly on the ratio $\xi_0\equiv m_0/m_{1/2}$. For $\xi_0\gg1$ the relic
abundance is almost always much too large, whereas for $\xi_0\ll1$ the opposite
occurs. For $\xi_0\sim1$ there are wide ranges of the remaining parameters for
which $\Omega_\chi\sim1$. We also determine that $m_{\tilde q}\gsim250\GeV$
and $m_{\tilde l}\gsim100\GeV$ are necessary in order to possibly achieve
$\Omega_\chi\sim1$. These lower bounds are much weaker than the corresponding
ones derived previously when radiative breaking was {\it not} enforced.
\bigskip
{\vbox{\baselineskip12pt\hbox{CERN-TH.6584/92}\hbox{CTP--TAMU--56/92}
\hbox{ACT--16/92}\hbox{UAHEP9212}}}
\Date{July, 1992}

\newsec{Introduction}
The fundamental observation by the COBE satellite of minute anisotropies in
the cosmic microwave background radiation \Smoot, has once again made evident
the synergism between particle physics and cosmology \dvn, since the most
compelling explanation for these observations \Wright\ appears to be found in
inflationary theories of the early universe \inflation. Theoretically, the
observed large-scale structure of the universe requires the presence of a
dominant cold dark matter (CDM) component in the energy density \CDM. However,
CDM may not be enough to produce the observed small-scale structure \NO, and
recent re-evaluations of this problem \many\ suggest the presence of a
sub-dominant hot dark matter (HDM) component as well, in the form of \
$\sim{\rm few}\eV$ neutrinos.

Even though the CDM component could have various origins \DM, from the particle
physics point of view it is customary to ascribe it to an ${\cal O}(100\GeV)$
stable particle, \ie, the lightest neutralino ($\chi$) (and also the lightest
supersymmetric particle) of the minimal supersymmetric Standard Model (MSSM)
\EHNOS. The standard calculational procedure consists of searching through the
vast parameter space of the model for `cosmologically favored' regions, that
is, regions with near critical relic abundance of neutralinos, \ie, where
$\Omega_\chi h^2_0\lsim h^2_0$, with $\Omega_\chi=\rho_\chi/\rho_0$ the
fraction of the present total energy density (assumed to be the critical
density) in the form of neutralinos, and
$h_0$ the Hubble parameter in units of $100{\rm\,km\,s^{-1}\,Mpc^{-1}}$
(current observations indicate that $0.5\le h_0\le1$). A more conservative
(\ie, cosmological-model--independent) outcome of this procedure is the
identification of regions of parameter space which are cosmologically
disfavored, \ie, regions where $\Omega_\chi h^2_0\gsim h^2_0$.

Since $\Omega_\chi$ depends on the annihilation cross section
$\chi\chi\to{\rm all}$, the whole set of couplings and masses in the model
need to be specified before $\Omega_\chi$ can be accurately computed. In the
MSSM twenty-one parameters are needed to specify the relevant quantities. To
make things tractable, several ad-hoc assumptions about the model parameters
are
usually made \refs{\EHNOS,\usualdm}, resulting in approximate results of some
interest but of limited scope. Things simplify dramatically when, following
well-established theoretical prejudices, the MSSM is embedded in a generic
supergravity model with universal soft supersymmetry breaking
\refs{\EZ,\Nojiri,\LNYa,\LNYb,\ER}, since then the parameter count drops to
just eight. The final step in this theoretical round-up is to enforce the
requirement of radiative electroweak symmetry breaking, which cuts down the
parameter count to five variables, introduces a correlation between the usual
relic density variables $(M_2,\mu)$, and determines the Higgs boson masses in
terms of the basic model parameters. (Hereafter we refer to this unified model
as the Supersymmetric Standard Model (SSM).)
The latter two novel effects have important consequences in the
determination of the allowed region of parameter space and in the value of
$\Omega_\chi h^2_0$ throughout this region. This ultimate calculation has been
performed before for a `no-scale'-inspired
supersymmetry breaking scenario (where $m_{1/2}\gg m_0,A\approx0$) imposing
the radiative breaking constraint at tree-level \noscale\ and one-loop \EZ,
although in both cases with the inaccurate approximation of tree-level
Higgs boson masses.\foot{Concurrently with our calculation, there has
appeared a new calculation \DNdm\ imposing radiative breaking at tree-level but
including the one-loop corrections to the Higgs boson masses, although in a
more restrictive supersymmetry breaking scenario where $A$ is determined,
and also assuming $\lambda_b(M_U)=\lambda_\tau (M_U)$.}
The purpose of this note is to present the results of this calculation in the
context of supergravity models with universal soft supersymmetry breaking and
radiative electroweak breaking in the one-loop approximation \aspects,
including the ensuing one-loop corrected Higgs boson masses \refs{\ERZ,\higgs}.

Besides the particle physics model used to calculate the
annihilation cross section, various approximations have been used regarding
the composition of the $\chi$ state, the dominant annihilation channels, and
the solution of the Boltzmann equation needed to determine $\Omega_\chi h^2_0$.
See Refs. \refs{\LNYa,\Griest,\DNdm} for a discussion of these matters.
The actual relic density calculations in this paper have been carried out
following the methods previously described in Ref. \LNYa.

\newsec{The favored regions}
The Supersymmetric Standard Model (SSM) is an $SU(3)\times SU(2)\times U(1)$
model with the minimal three generations and two Higgs doublets of matter
representations, and which is assumed to unify into a larger gauge group at a
unification mass of $M_U\approx10^{16}\GeV$. The parameter space of this model
can be described in terms of three universal soft-supersymmetry breaking
parameters: $m_{1/2},m_0,A$, the top-quark mass $m_t$, and the ratio of Higgs
vacuum expectation values $\tan\beta=v_2/v_1$; the sign of the superpotential
Higgs mixing term $\mu$ is also undetermined \noscale. Several consistency and
phenomenological constraints restrict the range of the model parameters
\refs{\noscale,\aspects}. For ease of comparison with earlier work
we calculate $\Omega_\chi h^2_0$ and plot it in the traditional relic density
$(M_2,\mu)$ plane for a given $\tan\beta$, where
$M_2=(\alpha_2/\alpha_U)m_{1/2}
\approx0.83 m_{1/2}\approx0.30 m_{\tilde g}$ is the $SU(2)$ gaugino mass. This
is accomplished by numerically inverting the relation $\mu=\mu(m_t)$. We also
take $\tan\beta=2,8,15,30$, $\xi_0\equiv m_0/m_{1/2}=0.0,0.2,0.4,0.6,0.8,1.0$
and $2,4,6,8$; $\xi_A\equiv A/m_{1/2}=-\xi_0,0,+\xi_0$; and vary
$0\le M_2\le300\GeV$ and $0\le|\mu|\le800\GeV$ so that $m_{\tilde g}\lsim1\TeV$
and fine-tuning of the parameters occurs at two-orders-of-magnitude or less
\aspects. The resulting set of figures gives a good sampling of the whole
parameter space. In Fig. 1 we present a representative set of plots for
$\Omega_\chi h^2_0$ in the $(M_2,\mu)$ plane for $\tan\beta=2,8$,
$\xi_0=0.0,0.4,1.0$, $\xi_A=0$, and both signs of $\mu$. Note that since we
have restricted $M_2<300\GeV$, then $m_\chi$ is also restricted: for large
$M_2$, $m_\chi<M_1={5\over3}\tan^2\theta_w M_2\approx{1\over2}M_2\lsim
150\GeV$.
This upper bound on $M_2$ also eliminates the possibility of nearly pure
higgsino $\chi$'s.\foot{For contours of $m_\chi$ and $\chi$-composition in the
$(M_2,\mu)$ plane see \eg, Fig. 1 in Ref. \LNYa.}

The first evident feature of these figures is that the
allowed parameter space is considerably constrained. For $\xi_0\sim1$ the
radiative breaking constraints exclude the lower diagonal portion of the
figures
due to the determination of $\mu$. For example, at tree-level \aspects,
\eqn\I{\mu^2_{tree}=m^2_{1/2}\left[X_{1/2}+\xi^2_0X_0
                        -\coeff{1}{2}(M_Z/m_{1/2})^2\right],}
and clearly for $\xi_0\lsim1$, $\mu\lsim m_{1/2}$ since $X_{1/2},X_0$ are
${\cal O}(1)$ functions \aspects. This correlation between $\mu$ and $M_2$ is
a distinctive feature of supergravity models with radiative electroweak
breaking and (for $\xi_0\lsim1$) eliminates a large portion of the otherwise
allowed parameter space (for example, compare Fig. 1c with Fig. 2 in
Ref. \LNYb\foot{We should point out that in Refs. \refs{\LNYa,\LNYb} $\mu$
followed an opposite sign convention than in the present paper and
Refs. \refs{\noscale,\higgs,\aspects}.}). Another consequence of the radiative
breaking constraints is the determination of the Higgs boson masses in terms
of the model parameters. This implies that as one moves around the $(M_2,\mu)$
plane in Fig. 1, the Higgs masses
are varying continuously. In fact, with growing $M_2$ (\ie, $m_{1/2}$), the
lightest Higgs boson ($h$) quickly approaches its upper bound (which is
variable
at one-loop but does not exceed $\approx150\GeV$ for $m_{\tilde g}<1\TeV$) and
the other three Higgs bosons become quite massive.
This means that: (i) the important Higgs-mediated $s$-channel annihilation into
fermion pairs becomes suppressed due to massive propagators as $M_2$ grows; and
(ii) the $\chi\chi\to hh$ channel is open throughout a fixed region in the
$(M_2,\mu)$ plane (\ie, where $m_\chi>m_h$). These two effects suppress the
annihilation rate and hence enhance the relic density relative to the usual
calculations (with \refs{\Nojiri,\LNYa,\LNYb,\ER} or without \usualdm\ the
supergravity mass relations) where the radiative breaking constraints are
not imposed and all the Higgs masses are kept fixed (at moderate values)
throughout the $(M_2,\mu)$ plane (for example, compare Fig. 1c with Fig. 2 in
Ref. \LNYb).

For lower values of $\xi_0$ the radiative breaking constraints apply as well
but the
phenomenological constraints bite more into the allowed region. The left edge
of the allowed area is determined primarily by the LEP lower bound on the
chargino mass ($m_{\chi^+}>45\GeV$) and the constraint $m_t\gsim90\GeV$. For
$\xi_0\approx0$ the allowed area is further suppressed \noscale\ by demanding
an electrically neutral lightest supersymmetric particle.

Within the allowed area the value of $\Omega_\chi h^2_0$ grows steadily with
$\xi_0$ since larger values of $m^2_{\tilde q,\tilde l}\approx
 m^2_{1/2}(c_i+\xi^2_0)$ suppress the important
$t$- and $u$-channel sfermion-mediated $\chi\chi\to f\bar f$ ($f=q,l$)
annihilation channels. On the other hand, $\xi_A$ does not affect the relic
density in any significant way since $A$ mainly determines the degree of
left-right stop mixing. However, $\xi_A$ affects the calculated value of $\mu$
(mainly for small $\mu$) \aspects, and therefore it affects the left edge of
the allowed area; $\xi_A>0$ ($\xi_A<0$) moves the left edge to the left
(right).

The effect of $M_2\approx0.83m_{1/2}\approx0.30m_{\tilde g}$ is to scale up
all Higgs and sparticle masses and therefore one would expect
$\Omega_\chi h^2_0$
to grow steadily with $m_{1/2}$. Even though this is generally the case, the
steady growth can be locally (in $m_{1/2}$) depleted due to the presence of
poles and thresholds of the annihilation cross section. The most prominent of
these are the $Z$- and $h$-poles ($\chi\chi\to Z,h\to f\bar f$) which occur
for $m_\chi\approx{1\over2}M_Z$ and $m_\chi\approx{1\over2}m_h$ respectively,
and the $\chi\chi\to hh$ threshold for $m_\chi\approx m_h$. In Fig. 2 (top
row) we show $\Omega_\chi h^2_0$ versus $m_{\tilde g}$ for $\xi_0=1,\xi_A=0$,
$m_t=150\GeV$, $\tan\beta=2,8$, and both signs of $\mu$ (solid $\mu>0$,
dashed $\mu<0$). These `rays' correspond
to a path of increasing $M_2$ and $|\mu|$ (and therefore $m_\chi$) through the
plots in Fig. 1c (with $M_2/|\mu|\approx0.5\,(0.7)$ for $\tan\beta=2\,(8)$)
such that $m_t=150\GeV$. For $\tan\beta=2$ and $\mu>0$ the $Z$-pole occurs for
$m_{\tilde g}\approx270\GeV$ and the $h$-pole does not occur (since $m_\chi>
{1\over2}m_h$ always). For $\mu<0$ the $h$- and $Z$-poles are close:
$m_\chi\approx{1\over2}m_h\approx37\GeV$ for $m_{\tilde g}\approx340\GeV$
and $m_\chi\approx{1\over2}M_Z$ for $m_{\tilde g}\approx380\GeV$. For
$\tan\beta=8$ the $h$- and $Z$-poles are even closer:
$m_\chi\approx{1\over2}M_Z$ for $m_{\tilde g}\approx310\,(370)\GeV$ and
$m_\chi\approx{1\over2}m_h\approx50\GeV$ for $m_{\tilde
g}\approx320\,(400)\GeV$
for $\mu>0$ ($\mu<0$). These features are evident in the figures. The
$\chi\chi\to hh$ threshold is also noticeable as a drop in the rate of increase
of $\Omega_\chi h^2_0$. The degree of effectiveness of this new annihilation
channel depends on the $\chi$-composition. Pure bino $\chi$'s do not couple to
the $h$ field. This is the case for the $\tan\beta=2$, $\mu>0$ ray where at
$m_{\tilde g}\approx440\GeV$, $m_\chi\approx m_h\approx70\GeV$ but only an
almost imperceptible drop occurs. The other rays encounter the threshold when
$\chi$ is of mixed composition and therefore the expected drop is clearly
observable. We remark that perhaps the most important effect of the
one-loop corrections to the Higgs boson masses is to shift the position of
the poles and thresholds discussed above, relative to the tree-level case.

Since $m_{1/2}$ and $\xi_0$ determine to a large extent the magnitude of
$\Omega_\chi h^2_0$, it is possible to scan the parameter space and determine
the lowest values of these parameters which give $\Omega_\chi h^2_0\sim h^2_0$.
We have done this for $h=1/2$ and find that it is necessary to have
$m_{\tilde q}\gsim250\GeV$ and $m_{\tilde l}\gsim100\GeV$ to possibly achieve
this. Lower values of $m_{\tilde q,\tilde l}$ do not suppress the annihilation
cross section sufficiently to give a large enough relic density. The bounds
are substantially weaker than previously obtained ({\it c.f.}
$m_{\tilde q}\gsim600\GeV$ and $m_{\tilde l}\gsim200\GeV$ \LNYa) without
imposing
the radiative breaking constraints since (as discussed above) now the important
Higgs-dependent annihilation channels also fade away with increasing $m_{1/2}$.

The effect of $\tan\beta$ on $\Omega_\chi h^2_0$ is felt in the
$\chi$-composition
for fixed $M_2$ and $\mu$, which affects the annihilation rate directly.
Heuristically we find that $\Omega_\chi h^2_0$ generally decreases for large
values of $\tan\beta$. This variable also affects the size of the allowed area
in the $(M_2,\mu)$ plane. The allowed area first increases with $\tan\beta$,
it peaks in size, and then it decreases for large $\tan\beta$ until it
disappears completely \aspects.

{}From Fig. 1 and the preceeding discussion it is clear that for $\xi_0\lsim1$
it is quite possible to obtain $\Omega_\chi h^2_0\sim h^2_0$, \ie, pluses,
squares, diamonds for $h=1,1/\sqrt{2},1/2$. As $\xi_0$ decreases the fraction
of the allowed area occupied by `preferred' points decreases and for
$\xi_0\lsim0.2$ it is no longer possible to obtain $\Omega_\chi h^2_0\sim
h^2_0$
for $0.5\le h_0\le 1$. However, $\xi_0\ll1$ may still provide enough neutralino
relic density for it to be the major component of the galactic halo.

\newsec{The excluded regions}
We have determined that for $\xi_0\lsim1$ and $m_{\tilde g}<1\TeV$ there are no
points in parameter space where $\Omega_\chi h^2_0>1$ and therefore one cannot
conclusively exclude any of these points on cosmological grounds. Note however
that if $h=1/\sqrt{2}\,(1/2)$ then $\Omega_\chi h^2_0>1/2\,(1/4)$ will become
disfavored. On the other hand, for $\xi_0\gsim2$ there are always regions of
parameter space where $\Omega_\chi h^2_0>1$ for almost all allowed values of
$M_2$. The fraction of the allowed area which becomes cosmologically excluded
increases with $\xi_0$.

To explore this dependence in a quantitative way it is better to change
variables from $(M_2,\mu)$ to $(M_2,m_t)$ for $\xi_0\gsim1$. This is because
for large $\xi_0$ values, the range of $\mu$ is much enlarged, \ie,
$\mu\lsim\xi_0 m_{1/2}$ (see Eqn. \I), whereas $m_t$ is always restricted to
be below $\approx190\GeV$ \refs{\noscale,\aspects}. We thus interchange $\mu$
with the basic parameter $m_t$. In Fig. 3 we show $\Omega_\chi h^2_0$ in the
$(M_2,m_t)$ plane for $\xi_0=2,\xi_A=0$, $\tan\beta=2,8$, and both signs of
$\mu$. Note the large number of cosmologically excluded points (at least the
stars).

The dependences of $\Omega_\chi h^2_0$ on the radiative breaking constraints,
and on $\xi_0,\xi_A,m_{1/2}$, and $\tan\beta$ are qualitatively as discussed
for the $\xi_0<1$ case above. The relic density increases steadily with
$\xi_0$,
except for values of $M_2$ where poles and/or thresholds occur. In Fig. 2
(bottom row) we show $\Omega_\chi h^2_0$ along a ray with $m_t=150\GeV$ in
Fig. 3. Compared to the top row, the bottom row has very similar features,
except for the overall scaling of the vertical axis. The reason for this
behaviour (\ie, the little change in $m_\chi$) is that
$\mu$ does not change too much for $\xi_0=1\to2$ and the $m_\chi$ contours
in the $(M_2,\mu)$ plane (see \eg, Fig. 1 in Ref. \LNYa) tend to change little
with $\mu$ (for large enough $\mu$); the change (and the corresponding shifts
in $m_\chi$) are larger for $\mu<0$. Note that in contrast with the
$\tan\beta=2,\mu>0$, $\xi_0=1$ case, for $\xi_0=2$ the $h$-pole does occur in
this case for $m_{\tilde g}\approx150\GeV$ and
$m_\chi\approx{1\over2}m_h\approx27\GeV$. Regarding the $\chi\chi\to hh$
thresholds in this case (see Fig. 2 bottom row), it is interesting to note that
for $\tan\beta=2$, $\mu>0$, the threshold effect, even though small it is
nevertheless noticeable (relative to the corresponding top row case in Fig. 2).
This is because the {\it relative} size of the drop in the annihilation cross
section due to the new channel is much larger here ($\xi_0=2$, small total
cross section) than in the previous case ($\xi_0=1$, large total cross
section).

To quantify the statement made above about larger values of $\xi_0$ giving
larger cosmologically excluded fractions of the allowed region, we define
$f_h\equiv n(\Omega_\chi h^2_0>h^2_0)/n_{tot}$, where
$n(\Omega_\chi h^2_0>h^2_0)$ is the number of points inside the allowed region
which have $\Omega_\chi h^2_0>h^2_0$ and $n_{tot}$ is the total number of
points for a fixed discrete gridding of the $(M_2,m_t)$ space
(with $\Delta m_t=\Delta M_2=5\GeV$). In Fig. 4 we plot $f_{h=1/2}$ versus
$\xi_0$ for $\xi_A=-\xi_0,0,+\xi_0$, $\tan\beta=2,8,15,30$, and both signs of
$\mu$. Due to the large extent of the computations to produce Fig. 4, these
were performed in the tree-level approximation to radiative breaking and the
Higgs boson masses (a full one-loop treatment would have required ${\cal O}
(\rm days)$ of NEC SX-3 supercomputer CPU time). However, in this particular
instance we do not expect the full one-loop treatment to modify significantly
the results in Fig. 4 (as verified explicitly in some cases) because its main
effect is to shift the positions of poles and thresholds, which basically do
not affect the value of $f_h$. As anticipated, as $\xi_0$ grows the fraction of
excluded points grows, and would probably approach $\approx100\%$ sooner were
it not for the poles. In fact, from Fig. 2 one can see that increasing $\xi_0$
reduces the size of the interval in $m_{\tilde g}$ where $\Omega_\chi h^2_0
<h^2_0$ ($=0.25$ in this case). Because these regions can only shrink but not
disappear completely, it is not possible to obtain an {\it absolute} upper
bound
on $\xi_0$ from cosmological considerations alone. However, large {\it
specific}
regions of parameter space can be ruled out in this way.

\newsec{Conclusions}
We have presented an exhaustive exploration of the parameter space of the
SSM which due to the constraints of universal soft supersymmetry breaking and
radiative breaking of the electroweak symmetry needs only five variables to
be described. The latter constraint imposes novel relations between the
traditional relic density variables and allows the determination of the Higgs
boson masses in terms of the basic parameters of the model. These two new
effects affect significantly the variation of $\Omega_\chi h^2_0$ throughout
the allowed parameter space. Furthermore, our calculations include the
state-of-the-art one-loop Higgs effective potential to enforce the radiative
breaking constraints and obtain automatically the one-loop corrected Higgs
boson masses. As such we believe this to be the ultimate calculational
framework upon which more specific supergravity models would need to be
investigated (see \eg, \troubles).

We find that $\Omega_\chi h^2_0$ depends most strongly on the parameter
$\xi_0=m_0/m_{1/2}$, as previously observed \LNYb. For $\xi_0\gg1$ the relic
density is almost always (barring accidental depletion due to $Z$- and/or
$h$-poles) in conflict with current cosmological observations. For $\xi_0\ll1$
the opposite occurs, although $\Omega_\chi h^2_0$ may still account for the
dark matter in the galactic halo. For $\xi_0\sim1$ there is a wide range of
the other parameters for which $\Omega_\chi\sim1$. Besides these qualitative
observations, one can draw more precise conclusions for specific points or
regions of parameter space. It is also possible to obtain an approximate
lower bound on the squark and slepton masses below which it is not possible
to get $\Omega_\chi\sim1$; we find $m_{\tilde q}\gsim250\GeV$ and $m_{\tilde l}
\gsim100\GeV$. These lower bounds are much weaker than those obtained
previously
\LNYa\ {\it without} enforcing the radiative breaking constraints.
Since on naturalness grounds \aspects\ we have limited $m_{\tilde g}$ to be
below 1 TeV, the value of $m_\chi$ is also restricted to $m_\chi\lsim150\GeV$.
These new bounds offer added hope that the supersymmetric spectrum may be
discovered at the next generation of $ee$ and $pp$ machines.

\bigskip
\bigskip
\bigskip
\noindent{\it Acknowledgments}: This work has been supported in part by DOE
grant DE-FG05-91-ER-40633. The work of J.L. has been supported in part by an
ICSC-World Laboratory Scholarship. The work of D.V.N. has been supported in
part by a grant from Conoco Inc. The work of K.Y. was supported in part by the
Texas National Laboratory Research Commission under Grant No. RCFY9155, and in
part by the U.S. Department of Energy under Grant No. DE-FG05-84ER40141. We
would like to thank the HARC Supercomputer Center for the use of their
NEC SX-3 supercomputer.
\listrefs
\listfigs
\bye